%% file: main.tex
\documentclass{article}
\usepackage{spconf,amsmath,graphicx}
\usepackage[acronym, nonumberlist]{glossaries}
\input{acronym}
\usepackage{amsmath,amssymb}
\usepackage{subcaption}
\usepackage{dsfont,comment,booktabs}
\usepackage{algorithm}
\usepackage{algorithmic}
\usepackage[hidelinks]{hyperref}
\usepackage{orcidlink}
\usepackage{xcolor}
\usepackage{placeins}

\title{Floating-Point Microformat Quantization and Pruning for Efficient MU-MIMO Neural Receivers}
\name{
  SaiKrishna Saketh Yellapragada$^{\star}$\orcidlink{0000-0002-7252-1580},
  Esa Ollila$^{\star}$\orcidlink{0000-0002-0898-5313},
  Mário Costa$^{\dagger}$\orcidlink{0009-0003-9488-2533},
  Yawei Li$^{\diamond}$\orcidlink{0000-0002-8948-7892}
  \thanks{Corresponding author: \texttt{saikrishna.yellapragada@aalto.fi}}
}
\address{
  $^{\star}$Department of Information and Communications Engineering, Aalto University, Finland \\
  $^{\dagger}$Nokia, Amadora, Portugal \\ $^{\diamond}$Nanyang Technological University, Singapore
}

\begin{document}
\ninept
\maketitle

\input{sections/00_Abstract}

\input{sections/01_Introduction.tex}
\input{sections/02_QAT_pruning_intro}
\input{sections/03_QATofNrx}
\input{sections/04_Experimental_Results}
\input{sections/06_Conclusion}

\clearpage
\subsection*{Acknowledgment}
The first author's work was supported by the European Union's 6GARROW project
(No.~101192194). The first and second authors thank Reinhard Wiesmayr
(ETH Zurich) for discussions on real-time neural receivers that improved
the clarity of this work. The first author thanks Sebastian Cammerer
(NVIDIA) for helpful correspondence. The authors thank the e-support team
at DICE, Aalto University, for GPU resources.

\bibliographystyle{IEEEbib}
\bibliography{references}

\end{document}

%% file: acronym.tex
\newacronym{6G}{6G}{Sixth Generation}
\newacronym{OFDM}{OFDM}{Orthogonal Frequency Division Multiplexing}
\newacronym{AI}{AI}{Artificial Intelligence}
\newacronym{DL}{DL}{Deep Learning}
\newacronym{BLER}{BLER}{Block Error Rate}
\newacronym{PTQ}{PTQ}{Post-Training Quantization}
\newacronym{QAT}{QAT}{Quantization-Aware Training}
\newacronym{SISO}{SISO}{Single-Input Single-Output}
\newacronym{SIMO}{SIMO}{Single-Input Multiple-Output}
\newacronym{MIMO}{MIMO}{Multiple-Input Multiple-Output}
\newacronym{LLR}{LLR}{Log-Likelihood Ratio}
\newacronym{LoS}{LoS}{Line-of-Sight}
\newacronym{NLoS}{NLoS}{Non-LoS}
\newacronym{LS}{LS}{Least-Squares}
\newacronym{LMMSE}{LMMSE}{Linear Minimum Mean Squared Error}
\newacronym{ZF}{ZF}{Zero-Forcing}
\newacronym{CPU}{CPU}{Central Processing Unit}
\newacronym{GPU}{GPU}{Graphics Processing Unit}
\newacronym{NPU}{NPU}{Neural Processing Unit}
\newacronym{RAN}{RAN}{Radio Access Networks}
\newacronym{BER}{BER}{Bit Error Rate}
\newacronym{FLOP}{FLOP}{Floating Point Operation}
\newacronym{CDL}{CDL}{Clustered Delay Line}
\newacronym{LDPC}{LDPC}{Low-Density Parity-Check}
\newacronym{RG}{RG}{Resource Grid}
\newacronym{DMRS}{DMRS}{Demodulation Reference Signal}
\newacronym{IFFT}{IFFT}{Inverse Fast Fourier Transform}
\newacronym{FFT}{FFT}{Fast Fourier Transform}
\newacronym{BCE}{BCE}{Binary Cross-Entropy}
\newacronym{CSI}{CSI}{Channel State Information}
\newacronym{SNR}{SNR}{Signal-to-Noise-Ratio}
\newacronym{STE}{STE}{Straight‑Through Estimator}
\newacronym{UE}{UE}{User Equipment}
\newacronym{V2X}{V2X}{Vehicle-to-Everything}
\newacronym{QAM}{QAM}{Quadrature Amplitude Modulation}
\newacronym{PHY}{PHY}{Physical Layer}
\newacronym{PUSCH}{PUSCH}{Physical Uplink Shared Channel}
\newacronym{MU}{MU}{Multi-User}
\newacronym{PRB}{PRB}{Physical Resource Block}
\newacronym{MCS}{MCS}{Modulation and Coding Scheme}
\newacronym{TDL}{TDL}{Tapped Delay Line}
\newacronym{CGNN}{CGNN}{Convolutional Graph Neural Network}
\newacronym{RE}{RE}{Resource Element}
\newacronym{UMi}{UMi}{Urban Microcell}
\newacronym{MLP}{MLP}{Multilayer Perceptron}
\newacronym{ReLU}{ReLU}{Rectified Linear Unit}
\newacronym{MAC}{MAC}{Multiply--Accumulate}
\newacronym{BOP}{BOP}{Bit Operation}

%% file: sections/00_Abstract.tex
\begin{abstract}
Neural receivers outperform conventional 5G~NR processing chains, but their
compute and memory demands hinder real-time deployment. For a
standard-compliant \gls{MU}-\gls{MIMO} neural receiver, the 4-bit number
\emph{format}, not merely the bit width, determines whether compression
preserves the gain over classical receivers. We apply weight and activation
\gls{QAT} and, separately, $50\%$ magnitude pruning, comparing
INT8/INT4 and FP8~(E4M3)/FP4~(E2M1) weights with INT8 post-\gls{ReLU}
activations. Trained on 3GPP UMi channels and evaluated on TDL-B and TDL-C,
8-bit weight--activation models remain within $0.05$\,dB of \texttt{FP32} at
$10\%$ and $1\%$ \gls{BLER}. At 4~bits, uniform INT4 loses
$3.3$--$3.7$\,dB and falls below LS--\gls{LMMSE}, whereas FP4 more than
halves this loss ($1.3$--$1.4$\,dB) and still outperforms it by about
$0.5$\,dB, even after pruning. FP4's denser near-zero grid matches the
trained weight distribution, and FP4 avoids the residual-path over-pruning
seen with INT4. An analytic cost model projects $66\times$ fewer
bit-operations and $8.8\times$ less weight storage for pruned 4-bit-weight
inference.
\end{abstract}

\begin{keywords}
Neural receivers, MU-MIMO, quantization-aware training, FP4, pruning, microformats
\end{keywords}

%
%

%% file: sections/01_Introduction.tex
\section{Introduction}
\label{sec:intro}
Neural receivers jointly learn channel estimation, equalization, and
demapping, improving \gls{BLER} over conventional pipelines
\cite{scamnrx,deeprx_mimo_icc}. Model-driven variants reduce
computation and parameters \cite{mahdi_mdx}. A standard-compliant
implementation already runs in real time \cite{wiesmayr2024design}.
Fine-tuning on 5G~NR measurements mitigates sim-to-real mismatch
\cite{nuri_reinhard_spawc}. Over-the-air tests confirm that
simulation-trained receivers transfer to single-antenna
software-defined radios, improving with site-matched training
\cite{riku_ota}. Deployment relies on \glspl{GPU} for real-time RAN
compute, yet a reported neural receiver requires
$\sim26$\,TFLOP/s, motivating quantization to reduce storage, memory traffic,
and bit-operation complexity \cite{ran_accelerated_compute_reinhard, park_simd}.

Prior integer weight-only \gls{QAT} for a \gls{SIMO} convolutional
receiver improves substantially over \gls{PTQ}, but uses uniform grids
and full-precision activations \cite{saketh_qat}. That study omits
post-\gls{ReLU} activation quantization, unstructured pruning, floating-point microformats, and standard-compliant
\gls{MU}-\gls{MIMO} reception. At the same nominal 4-bit weight
budget, OCP E2M1
(FP4) places reconstruction levels more densely near zero, whereas
INT4 uses uniform bins \cite{ocp2023mx}. We therefore compress the
standard-compliant receiver from prior work, whose layers combine a
graph neural network (GNN) and convolutional neural networks (CNNs)
\cite{scamnrx,wiesmayr2024design}. We test whether the FP4
advantage persists under INT8 activation quantization or global
pruning, and characterize residual-kernel sparsity.

Our contributions are threefold:
\begin{itemize}
  \item Adding INT8 post-\gls{ReLU} activation quantization to 8-bit
    weight-\gls{QAT} models degrades the required $E_b/N_0$ by at most
    $0.08$\,dB.
  \item At 4~bits, FP4~(E2M1) outperforms INT4 before and after $50\%$
    unstructured pruning.
  \item An analytic model quantifies weight storage and bit-operations.
\end{itemize}
Weight--activation \gls{QAT} and pruning are evaluated separately.

%% file: sections/02_QAT_pruning_intro.tex
\section{System Model and Neural Receiver}
\label{sec:sys_nrx}


\subsection{Uplink Setup}
\label{sec:system_model}
$U$ single-layer \glspl{UE} transmit over the same time--frequency
resources to an $N_{\mathrm{rx}}$-antenna base station. After OFDM demodulation, the
received vector at subcarrier $s$ and OFDM symbol $t$ is
\begin{equation}
\label{eq:mu_mimo}
  \mathbf{y}_{s,t}
  = \sum_{u=1}^{U}
    \mathbf{h}_{s,t,u}\,x_{s,t,u}
    + \mathbf{n}_{s,t},
  \qquad
  \mathbf{n}_{s,t}\sim\mathcal{CN}(\mathbf{0},N_0\mathbf{I}_{N_{\mathrm{rx}}}),
\end{equation}
where $u$ indexes the \glspl{UE},
$x_{s,t,u}\in\mathbb{C}$ is the symbol transmitted by UE $u$,
$\mathbf{h}_{s,t,u}\in\mathbb{C}^{N_{\mathrm{rx}}}$ is the corresponding channel vector, and
$\mathbf{y}_{s,t},\mathbf{n}_{s,t}\in\mathbb{C}^{N_{\mathrm{rx}}}$ are the received
and noise vectors, respectively. Here, $\mathcal{CN}$ denotes a
circularly symmetric complex Gaussian distribution, $N_0$ is the noise
variance per receive component, and $\mathbf{I}_{N_{\mathrm{rx}}}$ is the
$N_{\mathrm{rx}}$-dimensional identity matrix. Payload bits are \gls{LDPC}-encoded
and mapped to \gls{QAM} symbols; known \gls{DMRS} tones provide an
initial \gls{LS} channel estimate. The neural receiver processes the
post-\gls{FFT} data and pilot grid and outputs per-bit \glspl{LLR} for
conventional 5G~NR decoding.

\vspace{-2mm}

\subsection{Neural Receiver Architecture}
\label{sec:architecture}
We adopt the GNN--CNN architecture for MU-\gls{MIMO}
\gls{PUSCH} reception \cite{scamnrx,wiesmayr2024design,neural_rx}. Each resource element has 18
input features. An input CNN with
channel sequence $18{\to}128{\to}128{\to}56$ initializes a state of
dimension $d_s{=}56$. Each of the $N_{\mathrm{it}}{=}8$ unrolled
blocks applies a $56{\to}64{\to}56$ user-mixing \gls{MLP} and a
$114{\to}128{\to}128{\to}56$ Update CNN to exchange information
across users and the time--frequency grid, respectively. A dense
readout of width $128$ produces \glspl{LLR}. Each separable convolution
comprises a depthwise spatial filter and a pointwise channel mixer.
Pruning targets pointwise, aggregation, and non-readout dense kernels.
Depthwise and readout kernels remain dense
(Section~\ref{sec:pruning}). Shared weights accommodate varying user
counts and bandwidths, enabling training on $4$ \glspl{PRB} and
evaluation on a $132$-\gls{PRB} grid.

\vspace{-2mm}

\subsection{Loss and Training Schedule}
\label{sec:loss}
Training minimizes bitwise \gls{BCE} with an auxiliary
channel-estimation loss. At iteration $k$, for coded bit $b\in\{0,1\}$,
let $\hat{L}^{(k)}$ denote the predicted \gls{LLR}, with positive
values favoring $b=1$, and let $\mathbf{h}$ and $\hat{\mathbf{h}}^{(k)}$
denote the true and estimated channel vectors respectively. The objective is
\begin{align}
  \label{eq:total_loss}
    \mathcal{L}
    &= \sum_{k=1}^{N_{\mathrm{it}}}
       \left(
         \mathcal{L}_{\mathrm{BCE}}^{(k)}
         + \lambda\,\mathcal{L}_{\mathrm{MSE}}^{(k)}
       \right), \\
  \label{eq:bce}
    \mathcal{L}_{\mathrm{BCE}}^{(k)}
    &= -\mathbb{E}\!\left[
         b\log\sigma(\hat{L}^{(k)})
         +(1-b)\log\!\bigl(1-\sigma(\hat{L}^{(k)})\bigr)
       \right], \\
  \label{eq:mse}
    \mathcal{L}_{\mathrm{MSE}}^{(k)}
    &= \mathbb{E}\!\bigl[
         \|\hat{\mathbf{h}}^{(k)}-\mathbf{h}\|_{2}^{2}
       \bigr],
  \end{align}
where $\sigma$ is the logistic sigmoid, $\mathbb{E}$ averages over data
bits and channel realizations, and $\lambda>0$ weights the auxiliary
loss. The coefficient $\lambda$ is set to $0.02$ and $0.01$ in
successive training phases. Training uses the open-source
neural-receiver implementation in Sionna \cite{sionna,neural_rx}.
Each mini-batch contains $N_B{=}128$ samples with random active-user
counts, velocities in $[0,56]$\,m/s, and rate-adjusted $E_b/N_0$ drawn
from phase-specific ranges spanning approximately $0$--$15$\,dB. The
\texttt{FP32} baseline is trained with Adam quantized models
warm-start from that checkpoint \cite{kingma2014adam}.

%% file: sections/03_QATofNrx.tex
\section{Quantization-Aware Training and Pruning for MU-MIMO Neural Receivers}
\label{sec:nrx_qat}

\subsection{Weight Fake Quantization}
\label{sec:fake_quant}
Let $\mathbf{W}$ denote a kernel (convolutional or dense) and let
$\mathcal{G}\subset\mathbb{R}$ be a finite quantization grid with
maximum magnitude $g_{\max}=\max_{g\in\mathcal{G}}|g|$. We use
\emph{per-channel abs-max} scaling along the output-channel axis:
indices $c$ and $i$ denote an output channel and a coefficient within
that channel, respectively.
\begin{align}
  s_c &= \frac{\max_{i}\,|W_{i,c}|}{g_{\max}},
  \label{eq:absmax_scale}\\
  \widehat{W}_{i,c}
  &=s_c\cdot\mathrm{round}_{\mathcal{G}}\!\bigl(W_{i,c}/s_c\bigr),
  \label{eq:fake_quant}
\end{align}
where $\mathrm{round}_{\mathcal{G}}(\cdot)$ maps to the nearest
element of $\mathcal{G}$. 

Following prior weight-\gls{QAT}, the forward pass uses fake-quantized
weights, while an \gls{STE} updates a full-precision master copy
\cite{understanding_STE}.
Abs-max \gls{PTQ} quantizes a frozen \texttt{FP32} checkpoint once
\cite{saketh_asilomar25}. Here, \eqref{eq:absmax_scale} is recomputed
from current master weights every training step, allowing adaptation
to rounding and nonuniform floating-point reconstruction levels.
Biases use the corresponding weight format. For INT$b$,
$b\in\{4,8\}$, the symmetric narrow-range grid is
$\mathcal{G}_{\mathrm{INT}b}=\{-g_{\max},\ldots,g_{\max}\}$, where
$g_{\max}=2^{b-1}-1$.

\vspace{-2mm}

\subsection{Floating-Point Microformats: FP8 and FP4}
\label{sec:fp_micro}
Unlike uniform integer grids, low-bit floating-point formats allocate
resolution nonuniformly, with denser levels near zero and wider dynamic
range \cite{micikevicius2022fp8,rouhani2023microscaling}. We use the
OCP E4M3 and E2M1 element formats \cite{ocp2023mx}. FP8~(E4M3) assigns
one sign, four exponent, and three mantissa bits, with maximum finite
magnitude $g_{\max}=448$. FP4~(E2M1) assigns one sign, two exponent,
and one mantissa bit, yielding the fifteen-level grid
\begin{equation}
\label{eq:fp4_grid}
  \mathcal{G}_{\mathrm{E2M1}}
  = \{0,\pm0.5,\pm1,\pm1.5,\pm2,\pm3,\pm4,\pm6\},
\end{equation}
where $g_{\max}=6$. E2M1 and INT4 both provide nominal $8\times$
weight compression relative to \texttt{FP32}, but differ in grid
spacing. All formats use the same abs-max scaling
\eqref{eq:absmax_scale} and \gls{STE} during \gls{QAT}.

\vspace{-2mm}

\subsection{Activation Quantization}
\label{sec:act_quant}
All convolutional and dense kernels (depthwise, pointwise,
aggregation, and readout) use \eqref{eq:fake_quant} with the INT8,
INT4, FP8~(E4M3), or FP4~(E2M1) grids. For a
post-\gls{ReLU} tensor $\mathbf{a}$, nonnegativity permits all 256
unsigned INT8 codes over $[0,\max(\mathbf{a})]$. The per-tensor scale
and fake-quantized activation are
$\Delta_a=\max(\mathbf{a})/255$ and
$\widehat{\mathbf{a}}=\Delta_a\operatorname{round}(\mathbf{a}/\Delta_a)$,
respectively \cite{jacob2018quantization}. All-zero tensors use
$\Delta_a=1$. Network inputs, LLR outputs, and channel-estimation
outputs remain in \texttt{FP32}. Joint weight and activation fake quantization lets
training adapt to both quantization errors. W$b$A8 denotes $b$-bit
weights with INT8 post-\gls{ReLU} activations.

\vspace{-2mm}

\subsection{Unstructured Magnitude Pruning}
\label{sec:pruning}
After weight \gls{QAT} converges, global magnitude pruning targets
pointwise convolution, aggregation, and non-readout dense kernels.
Depthwise and readout kernels remain dense
\cite{song_han_1,kuzmin2023pruning}. Let $\mathcal{T}$ index eligible
coefficients, $W_i$ denote coefficient $i$, and $M_i$ its binary keep
mask. A global threshold $\tau$ sets sparsity $\rho$ through
\begin{equation}
\label{eq:prune_mask}
  M_i=\mathds{1}_{\{|W_i|\ge\tau\}},\qquad
  W_i^{\mathrm{pr}}=M_iW_i .
\end{equation}
The removed fraction is
$\rho=|\{i\in\mathcal{T}:|W_i|<\tau\}|/|\mathcal{T}|$. We set
$\rho=50\%$ and freeze $\mathbf{M}$ during recovery fine-tuning,
enforcing $\mathbf{W}\leftarrow\mathbf{M}\odot\mathbf{W}$ after each
update, where $\odot$ denotes element-wise multiplication.

\vspace{-2mm}

\subsection{Training Protocol}
Weight-only \gls{QAT} starts from the pretrained \texttt{FP32}
checkpoint \cite{neural_rx}. Each W$b$A8 model is fine-tuned from its
corresponding weight-only checkpoint with both fake quantizers active.
Separately, sparse models are formed from weight-only checkpoints and
undergo fixed-mask recovery fine-tuning. Table~\ref{tab:sim_params}
lists batch sizes, iteration budgets, and learning rates.

%% file: sections/04_Experimental_Results.tex
%

\section{Experimental Results and Discussion}
\label{sec:results}

\begin{table}[t]
\centering
\caption{System, training, and evaluation parameters.}
\label{tab:sim_params}
\footnotesize
\setlength{\tabcolsep}{3pt}
\begin{tabular}{@{}p{0.44\columnwidth}p{0.50\columnwidth}@{}}
\toprule
Parameter & Value \\
\midrule
Users $\times$ layers (eval) & $2\times1$ \\
Receive antennas $N_{\mathrm{rx}}$ & 4 \\
Carrier / SCS & 2.14\,GHz / 30\,kHz \\
OFDM symbols & 14 \\
PRBs (train / eval) & 4 / 132 \\
MCS / code rate $R$ (Table~1 \cite{3gpp38214})
  & index 14 / $R\approx0.48$ (16-QAM) \\
DMRS & type~1, mapping~A, +1 position \\
Channel (train) & 3GPP UMi, speeds $[0,56]$\,m/s \\
Channel (eval) & UE1 TDL-B100/400 ($100$\,ns, $400$\,Hz) \cite{scamnrx} \\
 & UE2 TDL-C300/100 ($300$\,ns, $100$\,Hz) \\
NRX depth $N_{\mathrm{it}}$ / state width $d_s$ & 8 / 56 \\
Batch size $N_B$ & 128 \\
FP8/FP4/INT8/INT4 QAT & $5{\times}10^{4}$ at $10^{-4}$; then $2{\times}10^{5}$ at $5{\times}10^{-5}$ \\
W$b$A8 fine-tune & $5{\times}10^{4}$ at $10^{-4}$; then $10^{5}$ at $5{\times}10^{-5}$ \\
Prune recovery & $5{\times}10^{3}$ at $5{\times}10^{-5}$; then $10^{4}$ at $2{\times}10^{-5}$ \\
\bottomrule
\end{tabular}\\[2pt]
{\scriptsize Optimizer rows report the step count and learning rate for two successive Adam phases.}
\end{table}

\begin{figure*}[t]
    \centering
    \begin{subfigure}[t]{0.48\linewidth}
        \centering
        \includegraphics[width=\linewidth]{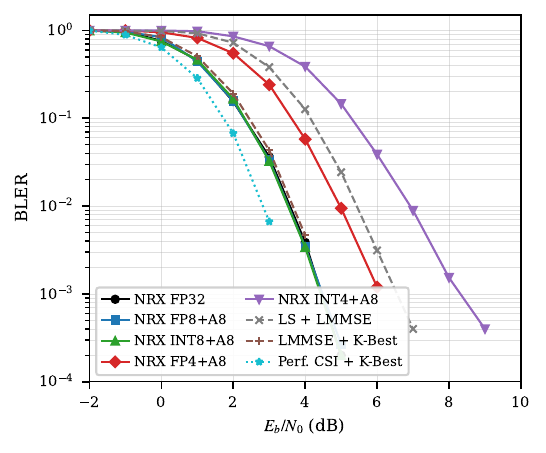}
        \caption{W$b$A8 QAT}
        \label{fig:bler_qat}
    \end{subfigure}\hfill
    \begin{subfigure}[t]{0.48\linewidth}
        \centering
        \includegraphics[width=\linewidth]{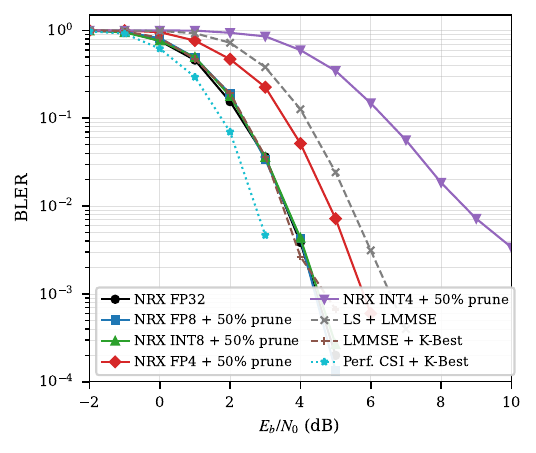}
        \caption{Weight QAT + 50\% pruning}
        \label{fig:bler_prune}
    \end{subfigure}
    \caption{\gls{BLER} versus $E_b/N_0$ on TDL-B100/400 \& TDL-C300/100 for
    (a) W$b$A8 \gls{QAT}, where $b\in\{4,8\}$ denotes the weight bit
    width, and (b) 50\%-pruned weight-\gls{QAT} checkpoints.}
    \label{fig:BLER}
\end{figure*}

\begin{figure*}[t]
    \centering
    \begin{subfigure}[t]{0.40\linewidth}
        \centering
        \includegraphics[width=\linewidth]{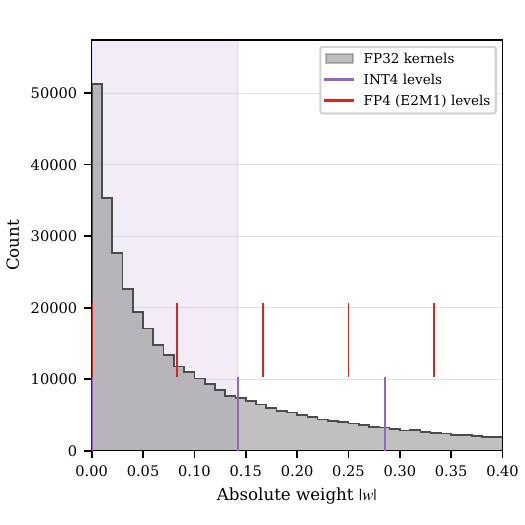}
        \caption{Normalized weight histogram.}
        \label{fig:weight_hist}
    \end{subfigure}\hfill
    \begin{subfigure}[t]{0.43\linewidth}
        \centering
        \includegraphics[width=\linewidth]{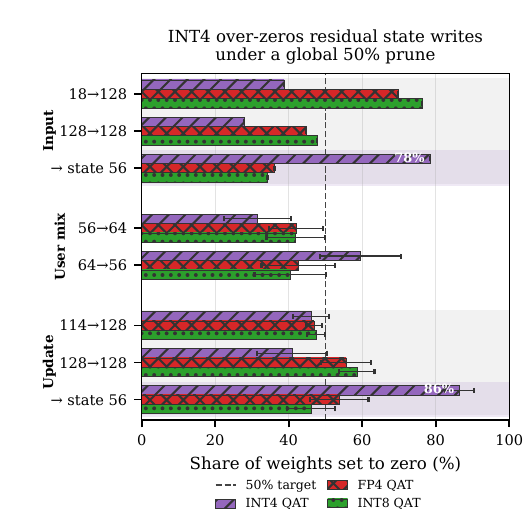}
        \caption{Per-kernel sparsity after global $50\%$ pruning.}
        \label{fig:sparsity_layers}
    \end{subfigure}
    \caption{FP4 robustness diagnostics. Panel~(a) compares
    per-channel abs-max-normalized \texttt{FP32} weight magnitudes
    with the INT4 and FP4 grids. Panel~(b) groups kernel sparsity by
    receiver stage after global $50\%$ pruning. Its y-axis labels
    give $C_{\mathrm{in}}{\to}C_{\mathrm{out}}$. Labels ending in
    ``state 56'' denote residual state writes. Error bars show the
    standard deviation across eight unrolled blocks. The Input CNN has
    one instance.}
    \label{fig:fp4_diagnostics}
\end{figure*}

\input{campaign_su_mu_20260809-080711/tables/mu_tdllow/wo_vs_a8}
\input{campaign_su_mu_20260809-080711/tables/mu_tdllow/complexity}

We evaluate weight \gls{QAT} with FP8~(E4M3), INT8, FP4~(E2M1), and
INT4. From each weight-only checkpoint, we separately derive a W$b$A8
model and a 50\%-pruned model. Experiments use the 3GPP-compliant
two-user PUSCH setup and optimizer budgets in
Table~\ref{tab:sim_params}.\footnote{Model weights and experiment scripts will be released at \url{https://github.com/saiksaketh/quantized_neural_rx/tree/feature/nrx-qat} upon acceptance to a possible IEEE conference venue.}
We report \gls{BLER} versus rate-adjusted $E_b/N_0$ at $10\%$ and
$1\%$ targets. Baselines include the full-precision (\texttt{FP32}) neural receiver,
LS channel estimation with \gls{LMMSE} equalization, \gls{LMMSE} channel
estimation with K-Best detection, and ideal K-Best detection with
perfect CSI (Fig.~\ref{fig:BLER}). An analytic model estimates storage
and bit-operations on the evaluation graph.

\vspace{-2mm}

\subsection{Activation Quantization is Radio-Neutral on Link Performance at 8~Bits}
\label{sec:res_a8}
Figure~\ref{fig:bler_qat} shows \gls{BLER} for the unpruned QAT
receivers with joint weight and activation quantization.
Table~\ref{tab:wo_vs_a8} reports the corresponding weight-only
operating points. FP8~(E4M3) and INT8 W$b$A8 receivers remain within
$0.05$\,dB of \texttt{FP32} at $10\%$ and $1\%$ \gls{BLER}. Adding A8
to the weight-only checkpoints changes the required $E_b/N_0$ by at
most $0.08$\,dB at either target. Both receivers outperform LS--\gls{LMMSE}
by approximately $1.83$--$1.91$\,dB and remain within
$\sim0.2$\,dB of \gls{LMMSE}--K-Best.\footnote{LS--\gls{LMMSE} and \gls{LMMSE}--K-Best use
floating-point arithmetic. Fixed-point arithmetic may increase
\gls{BLER} due to additional quantization error.} Thus, A8 causes no observed link
penalty in this setting. Section~\ref{sec:res_prune} evaluates pruning
from weight-only checkpoints.

\vspace{-2mm}

\subsection{FP4 Halves the 4-bit Loss}
\label{sec:res_fp4}
For weight-only QAT, the format strongly affects 4-bit performance
(Table~\ref{tab:wo_vs_a8}). At $10\%$ and $1\%$ \gls{BLER}, FP4~(E2M1)
needs $1.32$\,dB and $1.40$\,dB more $E_b/N_0$ than \texttt{FP32}
and beats LS--\gls{LMMSE} by about $0.5$\,dB. INT4 needs $3.3$--$3.7$\,dB
more than \texttt{FP32}. That gap is $1.98$\,dB and $2.27$\,dB larger
than FP4's, and INT4 also trails LS--\gls{LMMSE}. The nonuniform
E2M1 grid therefore more than halves the 4-bit SNR penalty at both
targets. Activation quantization retains FP4's advantage (Fig.~\ref{fig:bler_qat}).

We apply the per-channel abs-max normalization in
\eqref{eq:absmax_scale} to the $0.404$\,M trained \texttt{FP32}
prune-eligible pointwise and dense weights
(Fig.~\ref{fig:weight_hist}). For a normalized weight $w$, $65\%$ of
values satisfy $|w|<1/7\approx0.14$, the first nonzero INT4 magnitude.
FP4 adds a level at $1/12\approx0.08$. Moreover, $79\%$ of the weights
satisfy $|w|<0.25$. Over $[0,0.25]$, INT4 provides two magnitudes,
including zero, whereas FP4 provides four magnitudes
($0$, $0.08$, $0.17$, and $0.25$). The denser E2M1 grid better
resolves the distribution body and provides reconstruction levels
unavailable under uniform INT4.

\vspace{-2mm}

\subsection{Post-QAT Magnitude Pruning}
\label{sec:res_prune}
Figure~\ref{fig:bler_prune} shows weight-only checkpoints after 50\%
global pruning of eligible kernels. Depthwise and readout layers
remain dense. Penalties use each format's unpruned
weight-only parent, not panel~(a). At 8~bits, pruning changes the
required $E_b/N_0$ by at most $0.13$\,dB at either target. FP4 changes
by $-0.07$/$-0.10$\,dB after pruning and remains ahead of LS--\gls{LMMSE}.
In contrast, pruning adds $0.95$--$1.29$\,dB to INT4, which trails
LS--\gls{LMMSE} by more than $2$\,dB. \emph{Thus, FP4 tolerates 50\% pruning
with negligible radio impact, whereas INT4 with pruning degrades sharply.}

Figure~\ref{fig:sparsity_layers} reports the fraction of weights
zeroed in each labeled kernel. The global 50\% target applies over all
eligible coefficients, not per kernel. Under INT4, residual
update-to-state kernels average $86.5\%$ zeros (maximum $91.4\%$),
compared with $53.8\%$ (maximum $66.2\%$) under FP4. The Input CNN
state-write kernel reaches $78.5\%$ and $36.1\%$, respectively. Thus,
INT4 concentrates pruning in residual paths, while FP4 sparsity
remains comparable to INT8. Together with the denser near-zero FP4
grid in Fig.~\ref{fig:weight_hist}, this pattern is consistent with
FP4's higher pruning robustness.

Relative to \texttt{FP32}, the modeled W4A8+P configuration yields
$66\times$ and $8.8\times$ reductions in \glspl{BOP} and weight
footprint, respectively. W8A8 and W4A8 reduce \glspl{BOP} by
$16\times$ and $33\times$, respectively, and reduce stored weight
footprints by $3.8\times$ and $7.1\times$. The radio experiments
evaluate W$b$A8 and pruning separately, whereas the +P rows project
their combined analytic cost. Table~\ref{tab:complexity} reports
analytic costs on the evaluation graph with 132~\glspl{PRB}, 2~users,
and 8 iterations. Each slot requires $19.22$\,G multiply--accumulate
operations (MACs). Of these, $76\%$ arise from $1{\times}1$ pointwise
convolutions, the dominant GEMM-like kernels targeted by low-bit
quantization. We estimate \glspl{BOP} as
\[
  \mathrm{BOPs}
  = \sum_{\ell}\mathrm{MAC}_{\ell}\,b_{w,\ell}b_{a,\ell},
\]
where $b_{w,\ell}$ and $b_{a,\ell}$ are the weight and activation
widths of layer $\ell$. For eligible pruned kernels,
$\mathrm{MAC}_{\ell}$ is scaled by the retained-weight fraction.

%% file: campaign_su_mu_20260809-080711/tables/mu_tdllow/wo_vs_a8.tex
\begin{table}[t]
    \centering
    \caption{Weight-only QAT vs.\ W+A8 on TDL-B100/400 \& TDL-C300/100, interpolated from
    the same paired eval. W is the weight-QAT checkpoint; W+A8 is
    quantization-aware fine-tuning of that checkpoint with INT8 asymmetric
    post-\gls{ReLU} activations based on Table~\ref{tab:sim_params}.
    $\Delta$ is $E_b/N_0$ of W+A8 minus W.}
    \label{tab:wo_vs_a8}
    \begin{tabular}{lcccc}
    \toprule
     & \multicolumn{2}{c}{$E_b/N_0$ @ $10\%$ (dB)} & \multicolumn{2}{c}{$\Delta$ A8$-$W (dB)} \\
    \cmidrule(lr){2-3}\cmidrule(lr){4-5}
    Format & W-only & W+A8 & $10\%$ & $1\%$ \\
    \midrule
    FP8 (E4M3) & 2.29 & 2.30 & +0.00 & +0.01 \\
    INT8 & 2.24 & 2.31 & +0.08 & +0.01 \\
    FP4 (E2M1) & 3.62 & 3.62 & +0.00 & $-0.01$ \\
    INT4$^{\ddagger}$ & 5.60 & 5.59 & $-0.01$ & $-0.01$ \\
    \bottomrule
    \end{tabular}\\[2pt]
    \footnotesize{$^{\ddagger}$Negative $\Delta$ reflects quantization-aware fine-tuning of the weight-only parent, not a benefit from activation quantization.}
    \end{table}
    

%% file: campaign_su_mu_20260809-080711/tables/mu_tdllow/complexity.tex
\begin{table}[t]
\centering
\caption{Analytic inference cost per 5G~NR slot (132~PRB, 2~users,
8 iterations). W/A denotes weight/activation bit widths, and P denotes
pruning at $\rho=50\%$. Weight storage includes biases, scales, and the
sparse bitmask.}
\label{tab:complexity}
\footnotesize
\setlength{\tabcolsep}{3.5pt}
\begin{tabular}{@{}lcccc@{}}
\toprule
Config. & W/A & Spars. & Wt.\ (KiB)  & BOPs (T) \\
\midrule
FP32        & 32/32 & --   & 1708.5  & 19.7 \\
FP16        & 16/16 & --   &  862.1  &  4.9 \\
W8A8        & 8/8   & --   &  453.1  &  1.2 \\
W8A8+P      & 8/8   & 50\% &  310.9 &  0.7 \\
W4A8        & 4/8   & --   &  241.6 &  0.6 \\
W4A8+P      & 4/8   & 50\% &  194.1 &  0.3 \\
\bottomrule
\end{tabular}
\end{table}

%% file: sections/06_Conclusion.tex
\vspace{-2mm}
\section{Conclusion and Future Work}
\label{conclusion}
Eight-bit weight--activation \gls{QAT} preserves link performance and
reduces modeled feature-map traffic in a standard-compliant 5G~NR
MU-\gls{MIMO} neural receiver. FP4 outperforms \gls{LS}--\gls{LMMSE}, whereas INT4
performs poorly, supporting microscaling formats for neural receivers \cite{rouhani2023microscaling,ocp2023mx}. Analytically, the weight-activation quantization followed by pruning achieves $66\times$ and
$8.8\times$ reductions in bit-operations and weight storage,
respectively. This robustness aligns with FP4's denser near-zero grid
and lower residual-path sparsity.

Future work will pursue hardware-efficient 2:4 structured sparsity and native
MXFP4 block scaling. Distillation-augmented \gls{QAT} offers a
complementary route to further improve ultra-low-bit neural receiver link performance
\cite{yawei_qat_distillation}.